\documentclass[%
 amsmath,amssymb,
 aps,
prc,
]{revtex4-1}

\usepackage{amsfonts}
\usepackage{amsmath}
\usepackage{cases}
\usepackage{txfonts}
\usepackage{amssymb}
\usepackage{mathrsfs}
\usepackage{graphicx}
\usepackage{dcolumn}
\usepackage{bm}

\begin{document}

\preprint{}

\title{Energy and Centrality Dependence of Chemical Freeze-out Thermodynamics parameters}

\author{N. Yu}
\email[]{ning.yuchina@gmail.com}
\affiliation{Key Laboratory of Quark $\&$ Lepton Physics (MOE) and Institute of Particle Physics, Central China Normal University, Wuhan, 430079, China}
\author{F. Liu}
\affiliation{Key Laboratory of Quark $\&$ Lepton Physics (MOE) and Institute of Particle Physics, Central China Normal University, Wuhan, 430079, China}
\author{K. Wu}
\affiliation{College of Science, China Three Gorges University, Yichang, 443002, China}

\date{\today}

\begin{abstract}
Driven by the Beam Energy Scan (BES) program at the RHIC, researches and discussions on the QCD phase diagram have flourished recently. In order to provide a reference from microscopic transport models, we performed a systematic analysis, using a multiphase transport (AMPT) model for the particle yields and a statistical model (THERMUS) for the thermal fit, for Au+Au collisions at $\sqrt{s_{\text{NN}}}$=7.7-200 GeV. It is found that at a fixed collision centrality the chemical freeze-out parameter, temperature $T_{\text{ch}}$, increases with collision energy and somehow saturates at certain values of $T_{\text{ch}}$ in collisions near $\sqrt{s_{\text{NN}}}$=10 GeV, indicating the limiting temperature in hadronic interactions; meanwhile the baryon chemical potential $\mu_B$ decrease with the collision energy. The saturation temperature is also found to be dependent on partonic interaction. At a given collision energy, it is found that both $T_{\text{ch}}$ and $\mu_B$ decrease towards more peripheral collisions in the grand canonical approach. The energy and centrality dependence of other chemical freeze-out parameters, strangeness chemical potential $\mu_S$, strangeness undersaturation factor $\gamma_S$, and the volume of the fireball $V$ are also presented in this paper. The chemical potential ratio $\mu_s/\mu_B$ is also compared with lattice QCD calculation. The AMPT default model gives better descriptions on both the particle yields and the chemical freeze-out parameters than those from the AMPT string-melting model.  
\begin{description}
\item[PACS numbers]
\verb+25.75.Nq, 24.10.Lx, 24.10.Pa+
\end{description} 
\end{abstract}

\maketitle

The main goal of ultrarelativistic heavy-ion collisions is to create a new state of matter, the quark-gluon plasma (QGP) in laboratories. The phase structure of strong interaction, where quarks and gluons are deconfined, can be studied by quantum chromodynamics (QCD). After this strongly coupled QGP was observed at the Relativistic Heavy Ion Collider (RHIC)~\cite{Star1-1}, attempts are being made to vary the colliding beam energy and to research the thermodynamics properties of QCD matter expressed in terms of a $T-\mu_B$ phase diagram, which lies at the heart of what the RHIC Beam Energy Scan (BES) program is all about~\cite{Stephanov-1,Stephanov-2,Mohanty}. 

The chemical freeze-out parameters describing thermodynamic properties of the QCD phase diagram can be extracted from statistical thermal model. It is a surprising success that this model can reproduce essential features of particle production in nucleus-nucleus collisions~\cite{Cleymans1,Becattini1}, suggesting that statistical production is a general property of the hadronization process. Chemical freeze-out is typically supposed to happen when inelastic scattering stops, and the particle identities are set until they decay~\cite{Vogt}. Previous fits to the experimental data showed that from SPS energies upwards the extracted $T_{\text{ch}}$ are very close to the cross-over temperature $T_{C}$,  predicted from lattice QCD of 170$-$195 MeV ~\cite{Fodor,Bazavov1} for a phase transition. This is one of the indications that QGP is formed in the heavy-ion collision. 

In order to provide a reference of the QCD phase diagram from the microscopic transport model, it is the subject of present paper to follow this idea and analyze the collision energy and centrality dependence of chemical freeze-out parameters, which can describe hadron multiplicities statistically. These will provide some information about the phase transition and the effects of the size of the excited strong interaction system. For this reason, we concentrate our effort on the analysis of particle yields obtained from a multiphase transport (AMPT) model of two classes, default (D) and string-melting (SM) at different energies and centralities in heavy-ion collisions. 

The AMPT is a hybrid model which consists of four main components: the initial conditions, partonic interactions, conversion from partonic to hadronic matter, and hadronic interactions. The initial conditions are generated by the heavy-ion jet interaction generator (HIJING) model~\cite{HIJING}. Zhang's parton cascade (ZPC) model~\cite{ZPC} is used for describing the scattering among partons, which includes only two-body scattering with cross sections obtained from the pQCD with screening masses. By changing the value of the screening mass, different cross sections can be obtained, which will be used in studying the effect of parton cross sections in heavy-ion collisions. Because there are only minijet partons in parton cascade, the partonic stage in the D model does not play any role for most final hadronic observables, and reasonable variations of parton cross section do not change these observables. Consequently, the hadronic degree of freedom dominates in the collision. In the SM model, all the strings are converted to partons. Therefore, the partonic interaction dominates in the collision, and observables depend on the parton cross section. In this paper, our results are presented by varying the parton cross section within 3 to 10 mb. In the hadronization process, partons are recombined with their parent strings when they stop interaction, and the resulting strings are converted to hadrons by the Lund string fragmentation model~\cite{Lund} in the D model. In the SM model, a quark coalescence model is used to combine partons into hadrons. The hadronic interactions are based on a relativistic transport (ART) model~\cite{ART}. In our study, about two million events for each configuration (different AMPT models and parton cross sections) and each collision energy were generated for the Au+Au collision. The termination time of hadronic cascade $t_{\text{hc}}$ is varied from 0.6 to 30 fm/c to study the effect of hadronic interactions on the chemical freeze-out parameters. Hadronic productions at mid-rapidity, i.e. $\left|y\right|\leqslant 0.1$, particularly the yields of $\pi^\pm$, $K^\pm$, $p$, $\bar{p}$, $\phi$, $\Lambda$, $\bar{\Lambda}$ and multistrange hyperons $\Xi^-$, $\bar{\Xi}^+$, $\Omega^-$, $\bar{\Omega}^+$ are extracted from the AMPT model for different collision energies and centralities. The definition of centrality is determined by the per-event charged particle multiplicity $N_{\text{ch}}$ for pseudorapidity range $\left| \eta\right|\leqslant 0.5$. 

The chemical properties of the bulk particle production can be addressed by statistical thermal model. It is assumed that the particle abundance of species $i$ can be parameterized by
\begin{equation}
\frac{N_i}{V}=g_i \int{\frac{d^3p}{\left(2\pi\right)^3}\left[ \gamma_s^{-\left|S_i^{'}\right|}\exp \left( {\frac{{{E_i} - {\mu _i}}}{{{T_{ch}}}}} \right)\pm 1\right]^{-1}}
\end{equation}
where
\begin{equation}
\mu_i=\mu_BB_i+\mu_QQ_i+\mu_SS_i 
\end{equation}
and $g_i$ is the spin-isospin degeneracy factor; $ T_\text{ch} $ is the chemical freeze-out temperature;  $ B_i $, $ S_i $, $ Q_i $ are the baryon number, strangeness, and charge, respectively, of hadron species $i$; $ \mu_B $, $ \mu_S $, and $ \mu_Q $ are the corresponding chemical potentials for these conserved quantum numbers; $ E_i $ is the energy of the particle; $ \gamma_s $ is the strangeness undersaturation factor; and $S_i^{'}$ is the number of valence strange and anti-strange quarks in particle $i$.

The code THERMUS~\cite{THERMUS} is utilized to perform a thermal fit to the particle yields from the AMPT model. Within the model, there is a freedom regarding the ensemble with which to treat conserved numbers $B$, $S$, and $Q$ in strong interactions. The chemical potentials for each of these quantum numbers allow fluctuations about conserved averages, which is a reasonable approximation only when the number of particles carrying the quantum number concerned is large. Three ensembles can be used in the model. Those are the grand-canonical ensemble (GCE), canonical ensemble (CE), and mix-strangeness canonical ensemble (SCE). The GCE is the most widely used in the application to heavy-ion collisions. Conservation laws for energy and quantum or particle numbers are enforced on average through the temperature and chemical potentials in this ensemble. The CE is applied in high energy elementary collision, such as $p+p$, $p+\bar{p}$, and $e^+e^-$collisions, in which quantum numbers are conserved exactly. However, at the low energy of SIS (GSI SchwerIonen(Heavy ion) Synchrotron) heavy ion collisions, low strange particles production requires a canonical treatment of $S$ where SCE is useful. GCE and SCE are used in our thermal fits. The parameters considered in these thermal fits are the freeze-out temperature $T_{\text{ch}}$, the baryon chemical potential $\mu_B$, the charge chemical potential $\mu_Q$, the strangeness undersaturation factor $\gamma_s$ and the radius of the fireball (the volume of fireball $V=4/3\pi R^3$). In the GCE fit, the strangeness chemical potential $\mu_S$ is likewise being considered. The canonical or correlation radius $R_C$ inside which strangeness was exactly conserved was set to be equal to $R$ in the SCE fit. THERMUS allows the assignment of separate decay chains to each input. In this way, the model is able to match the specific feed-down corrections of a particular data set. The particle yields extracted from the AMPT model are corrected from weak decay feed-down. The analysis is performed by searching for the minimum of
$\chi^2$, that is,
\begin{equation}
\chi^2=\sum_i\frac{(N_i^{\text{exp}}-N_i^{\text{theo}})^2}{\sigma_i^2}
\end{equation}
in which $N_i^{\text{exp}}$ and $N_i^{\text{theo}}$ are the yields of the $i$th hadron species from the AMPT and from THERMUS, respectively. $\sigma_i$ is the statistical error from the AMPT model.

\begin{figure}[ht]
\scalebox{0.7}[0.7]{\includegraphics{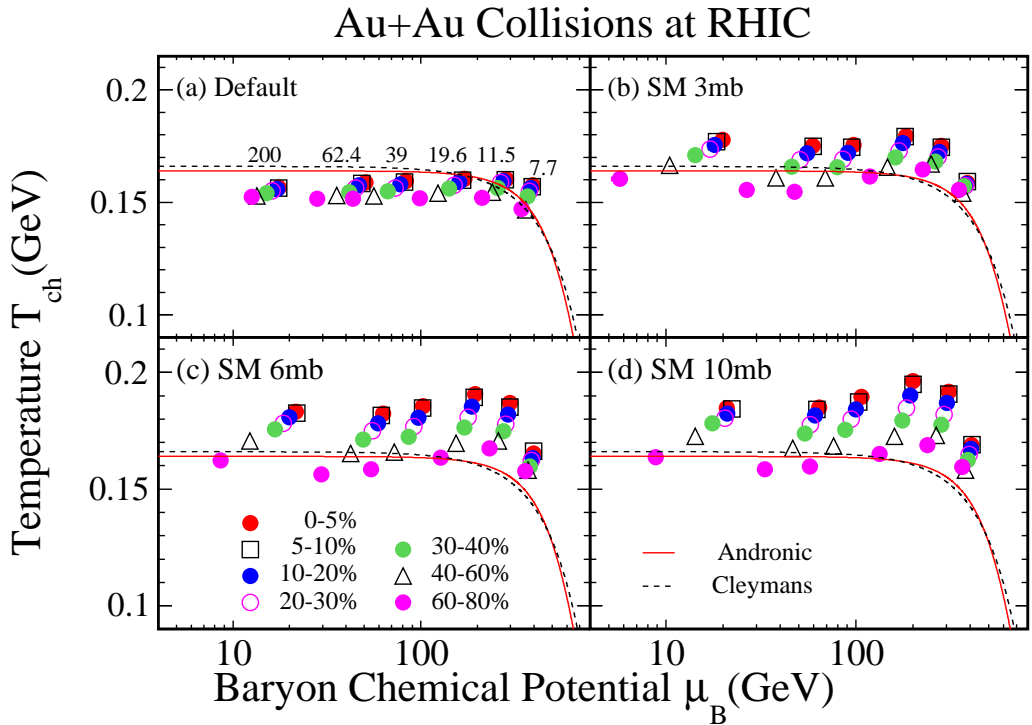}}
\caption{\label{tmub} Chemical freeze-out parameters temperature $T_{\text{ch}}$ versus baryon chemical potential $\mu_B$ from Au+Au collisions at $\sqrt{s_{NN}}=$7.7-200 GeV. The particle yields are from different AMPT models for $t_{hc}=$30 fm/c. The symbols are drawn from the THERMUS with GCE approach. The results from the D model are shown in the plot (a). Plots (b)-(d) are from the SM models with various partonic interaction cross sections. The curves are phenomenological parametrizations to quantity the experimental data from SPS to RHIC energies for the top 0\%$-$5\% centrality collisions~\cite{Andronic,Cleymans2}.}
\end{figure}

Figure~(\ref{tmub}) shows the energy and centrality dependence of the chemical freeze-out parameters $T_{\text{ch}}$ and $\mu_B$ from Au+Au collisions at $\sqrt{s_{NN}}=$7.7, 11.5, 17.3, 19.6, 27, 39, 62.4, and 200 GeV (the symbols from 17.3 and 27 GeV were removed for a clearer view). These symbols were extracted from the THERMUS GCE fit to the particle yields from the AMPT model for $t_{hc}=$ 30 fm/c. The solid and dashed curves are two numerical parametrizations of $T_{\text{ch}}$ and $\mu_B$ presented in Ref.~\cite{Andronic} and Ref.~\cite{Cleymans2}, respectively. The results from the D model are presented in the plot (a). Plots (b) -(d) are from the SM models with various partonic interaction cross sections. In the D model, at a fixed collision centrality, $T_{\text{ch}}$ shows weak dependence on the collision energy ($T_{ch}\approx$160MeV in the top 5$\%$) and $\mu_B$ decrease with increasing energy. In the SM model, $T_{\text{ch}}$ increases from low to high energy and somehow saturate for some centralities at a certain values ($T_{\text{ch}}$=170$-$190 MeV in the top 5$\%$) for $\sqrt{s_{NN}}\approx$10 GeV. The value of saturation temperature is found to be dependent on the parton cross section. This result indicates that the partonic interaction can enlarge $T_{\text{ch}}$ and not make significant changes in the value of $\mu_B$. 

\begin{figure}[ht]
\scalebox{0.7}[0.7]{\includegraphics{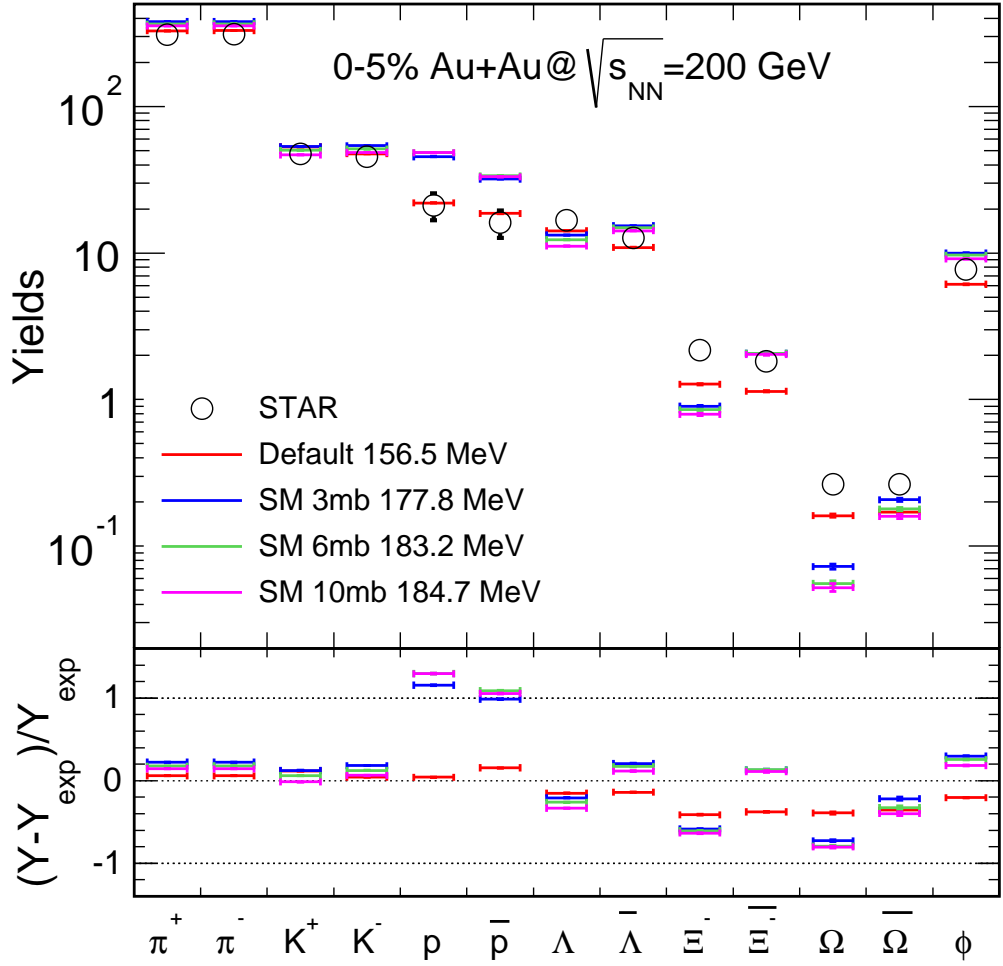}}
\caption{\label{pyield} Upper panel:Particle yields from the STAR experiment~\cite{Star3,Star4,Star5} and different AMPT models at $\sqrt{s_{NN}}=$200 GeV for the top 5$\%$ central Au+Au collision. $T_{\text{ch}}$'s are labeled in the legend.  Lower panel: Ratios of particle yield difference between the AMPT and the STAR data relative to the STAR data.}
\end{figure}

The particle yields from the STAR experiment and some different AMPT models, and ratios of the particle-yield difference between the AMPT and the STAR data relative to the STAR data, are shown at $\sqrt{s_{NN}}=$200 GeV for the top 5$\%$ central collisions in Fig.~\ref{pyield}. It's found that the D model gives a reasonable description of particle yields. For $\pi$ and $K$, the difference between these models is quite small. However, the yields of the proton and anti-proton extracted from the SM model are nearly twice as large as those from the D model. $T_{\text{ch}}$ can be determined by the particle/$\pi$ ratios which read, in the Boltzmann approximation, neglecting width and quantum statistics for a tentative consideration of generic features as follows~\cite{Cleymans3}:

\begin{equation}
\frac{N_i}{N_\pi}=\gamma_s^{\left|S_i^{'}\right|}\frac{g_i}{g_\pi}\frac{m_i^2}{m_\pi^2}\frac{K_2(m_i/T_{ch})}{K_2(m_\pi/T_{ch})}\frac{\text{exp}(\mu_i/T_{ch})}{\text{exp}(\mu_\pi/T_{ch})}
\end{equation}
where $\text{exp}(\mu_i/T_{ch})$ can be fixed by the particle/antiparticle ratios.
\begin{equation}
\frac{N_i}{N_{\bar{i}}}=\text{exp}(2\mu_i/T_{ch})
\end{equation}
Then the relation between particle/$\pi$ ratios and $T_{\text{ch}}$ can be written as
\begin{equation}
\sqrt{\frac{N_iN_{\bar{i}}}{N_\pi^+N_\pi^-}}\propto\frac{K_2(m_i/T_{ch})}{K_2(m_\pi/T_{ch})}
\end{equation}
Lowering the temperature $T_{\text{ch}}$ has the effect of bringing the particle/$\pi$ ratios down~\cite{Cleymans3}. The ratios of proton/$\pi$ simulated by the SM model are greater than those by the D model, which leads to larger $T_{\text{ch}}$ from the SM model. In experiment, due to the relatively large multiplicity of proton and anti-proton to those of strange baryons, high-precision proton measurement and feed-down correction are necessary in order to extract $T_{\text{ch}}$ and to map the chemical freeze-out line of the QCD phase diagram. 

\begin{figure}[ht]
\scalebox{0.7}[0.7]{\includegraphics{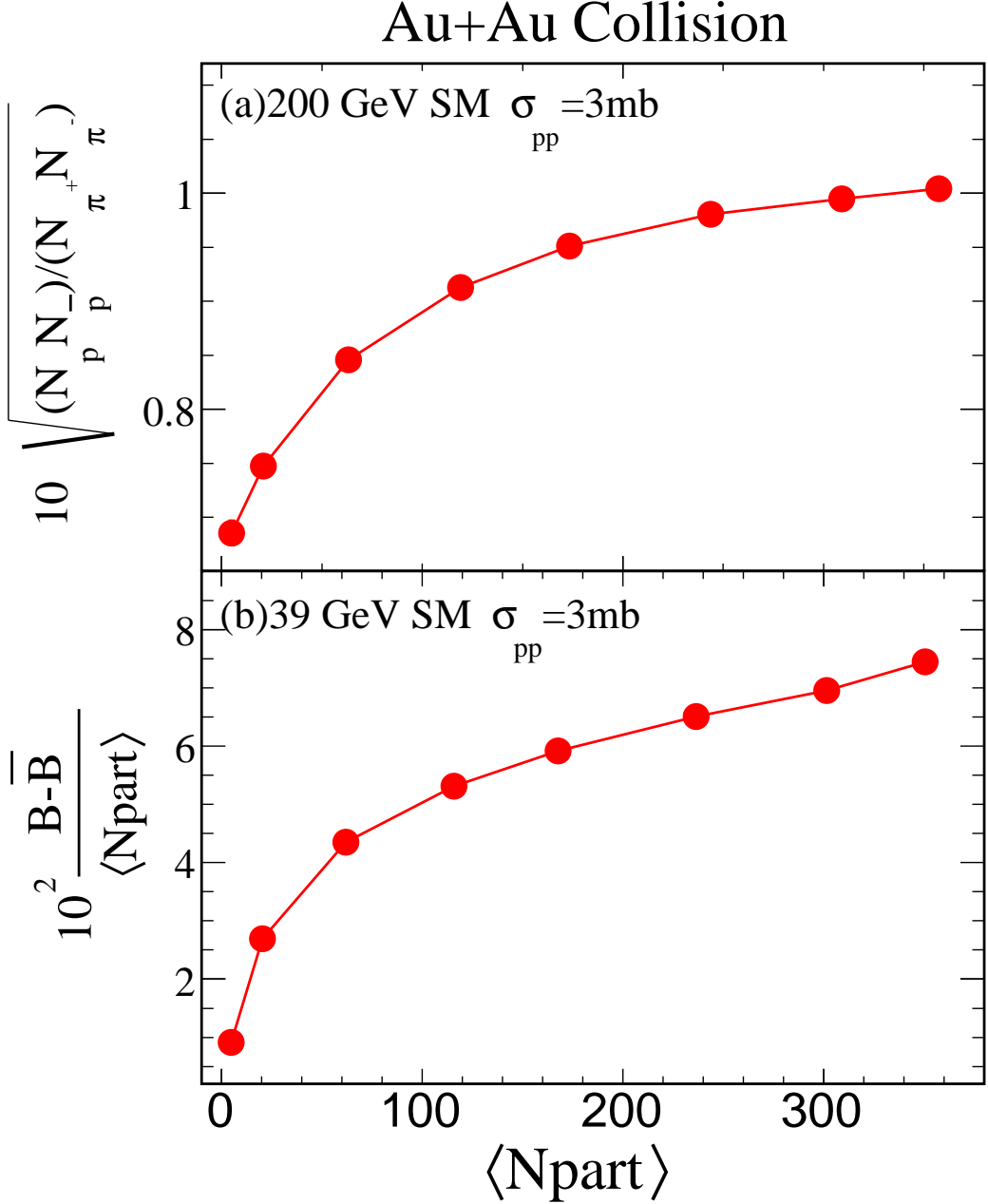}}
\caption{\label{pratios} Top Panel:Proton-to-$\pi$ ratios from the SM model with $\sigma_{pp}$=3 mb at $\sqrt{s_{NN}}=$200 GeV Au+Au collision. Bottom Panel:Net baryons (include proton, $\Lambda$, $\Xi$, $\Omega$ and their antiparticles) over $N_{\text{part}}$ as function as $N_{\text{part}}$ from the SM model with $\sigma_{pp}$=3 mb at $\sqrt{s_{NN}}=$39 GeV Au+Au collision.}
\end{figure}

\begin{figure}[ht]
\scalebox{0.7}[0.7]{\includegraphics{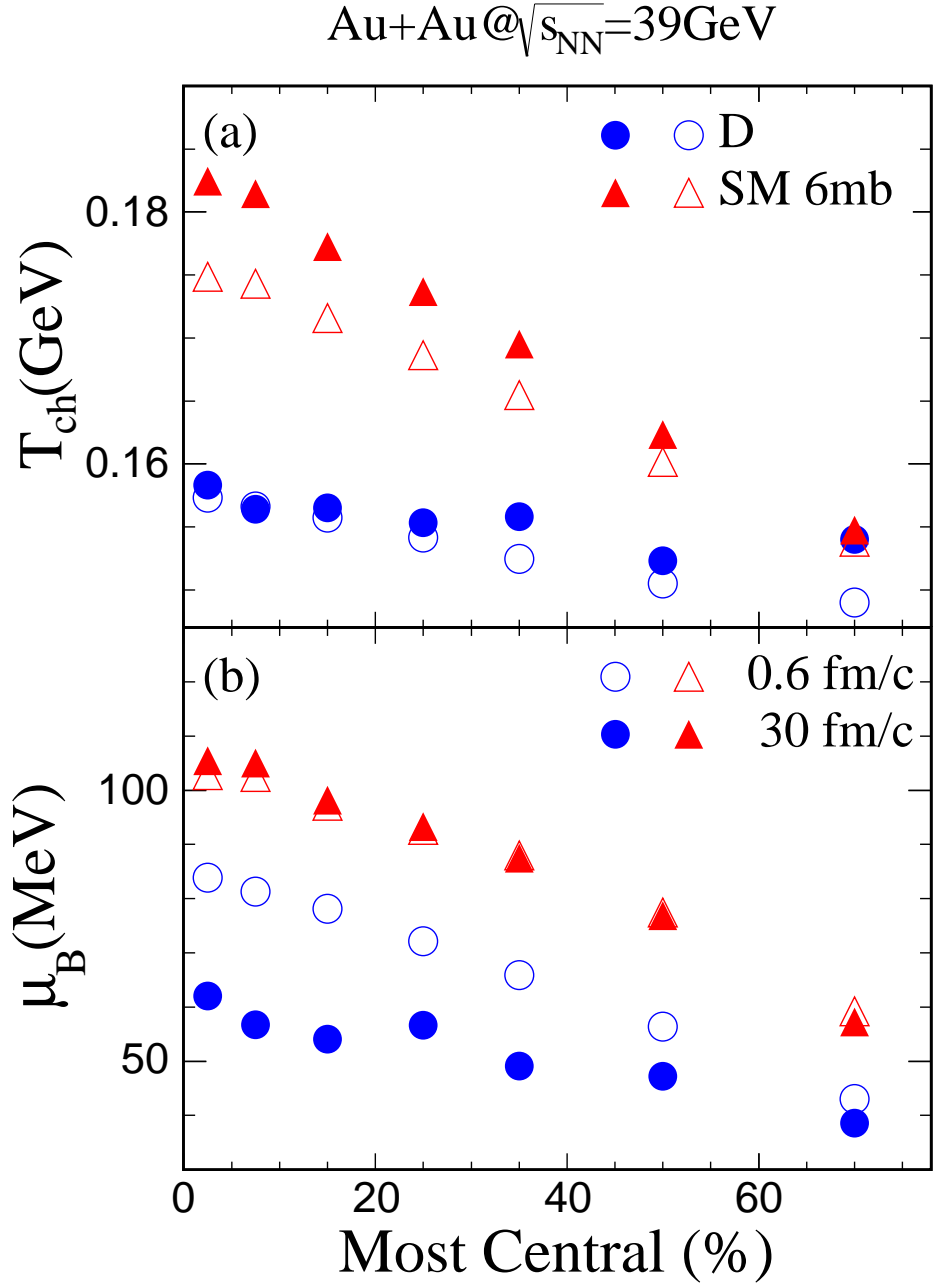}}
\caption{\label{harcas} Chemical freeze-out temperature $T_{\text{ch}}$ and $\mu_B$ as functions of centrality, with the termination time of hadron cascade $t_{\text{hc}}$=0.6 (open symbols) and 30 fm/c (filled symbols) from the D model (circles) and the SM model (triangles) with $\sigma_{pp}$=3 mb at $\sqrt{s_{NN}}=$39 GeV Au+Au collision.}
\end{figure}

At a given collision energy, it is found that both $T_{\text{ch}}$ and $\mu_B$ decrease from central to peripheral collisions in the D and the SM models. The top panel of figure~(\ref{pratios}) shows the proton/$\pi$ ratio as a function as collision centrality from the SM model with $\sigma_{pp}$=3 mb at $\sqrt{s_{NN}}=$200 GeV. This ratios decrease from central to peripheral, which is in agreement with the behavior of $T_{\text{ch}}$. The baryon chemical potential $\mu_B$ is related to the net-baryon density. In the bottom panel of Fig.~\ref{pratios}, we show the net baryons (including $p$, $\Lambda$, $\Xi$, $\Omega$ and their antiparticles) over $N_{\text{part}}$ (the number of participant nucleons) as a function as $N_{\text{part}}$ from the SM model with $\sigma_{pp}$=3 mb at $\sqrt{s_{NN}}=$39 GeV, which can describe the net-baryon density because $N_{\text{part}}$ is almost linear with system volume. It can be found that this value decreases from central to peripheral which leads the behavior of $\mu_B$. The results from the SCE fit are similar to the results above. In our event-by-event simulation shown in Fig.~\ref{harcas}, it is found that decreasing the hadronic rescattering by decreasing $t_{\text{hc}}$ to 0.6 fm/c does not change the trends of energy and centrality dependence but leads to a stronger centrality dependence of $T_{\text{ch}}$ in the SM model and a stronger centrality dependence of $\mu_{B}$ in the D model. From the top panel of Fig.~\ref{harcas}, the difference of $T_{\text{ch}}$ between $t_{\text{hc}}$=0.6 and 30 fm/c in central collisions is larger than that in peripheral collisions, which may hint that the system in central collision is not in chemical freeze-out status in the SM model.

\begin{figure}[ht]
\scalebox{0.7}[0.7]{\includegraphics{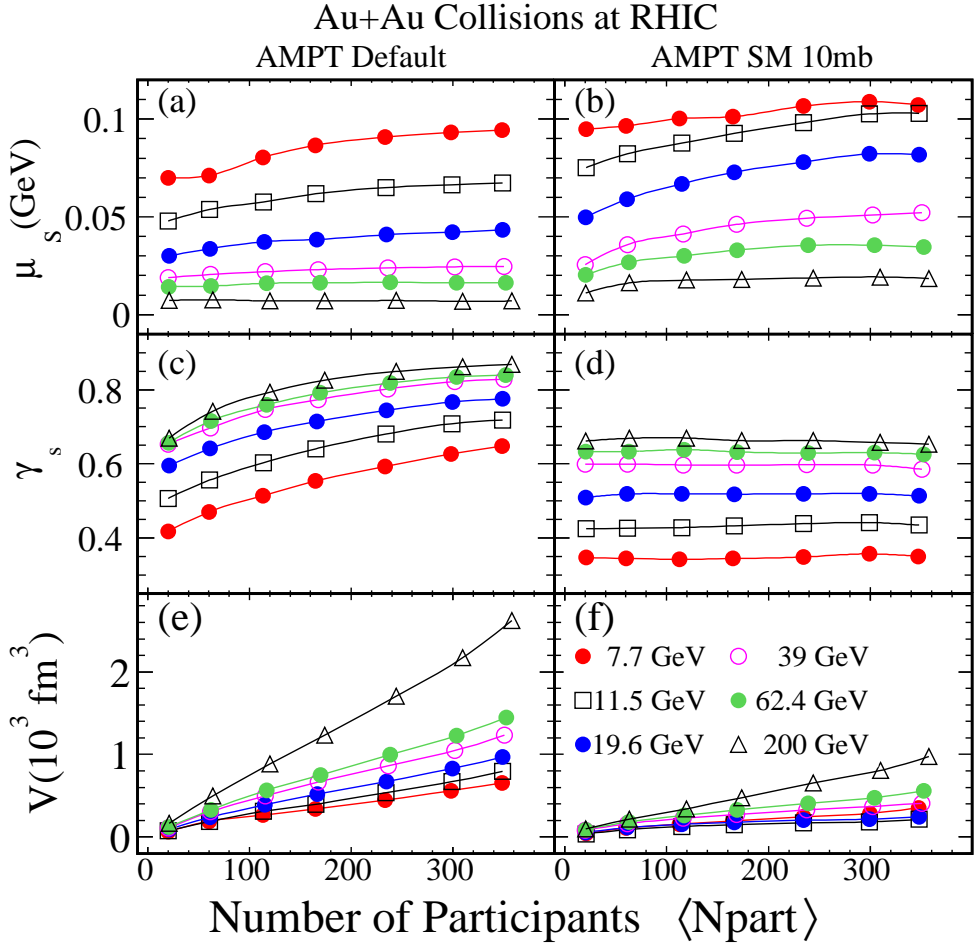}}
\caption{\label{otp} Chemical freeze-out parameters, strangeness chemical potential $\mu_S$, strangeness undersaturation factor $\gamma_s$, and the volume of the fireball $V$ of chemical freeze-out from the THERMUS GCE fit as functions as $N_{\text{part}}$. Left panels are the results from the D model; right panels are the results from the SM model for $\sigma_{pp}$=10 mb.}
\end{figure}

\begin{figure}[!ht]
\scalebox{0.7}[0.7]{\includegraphics{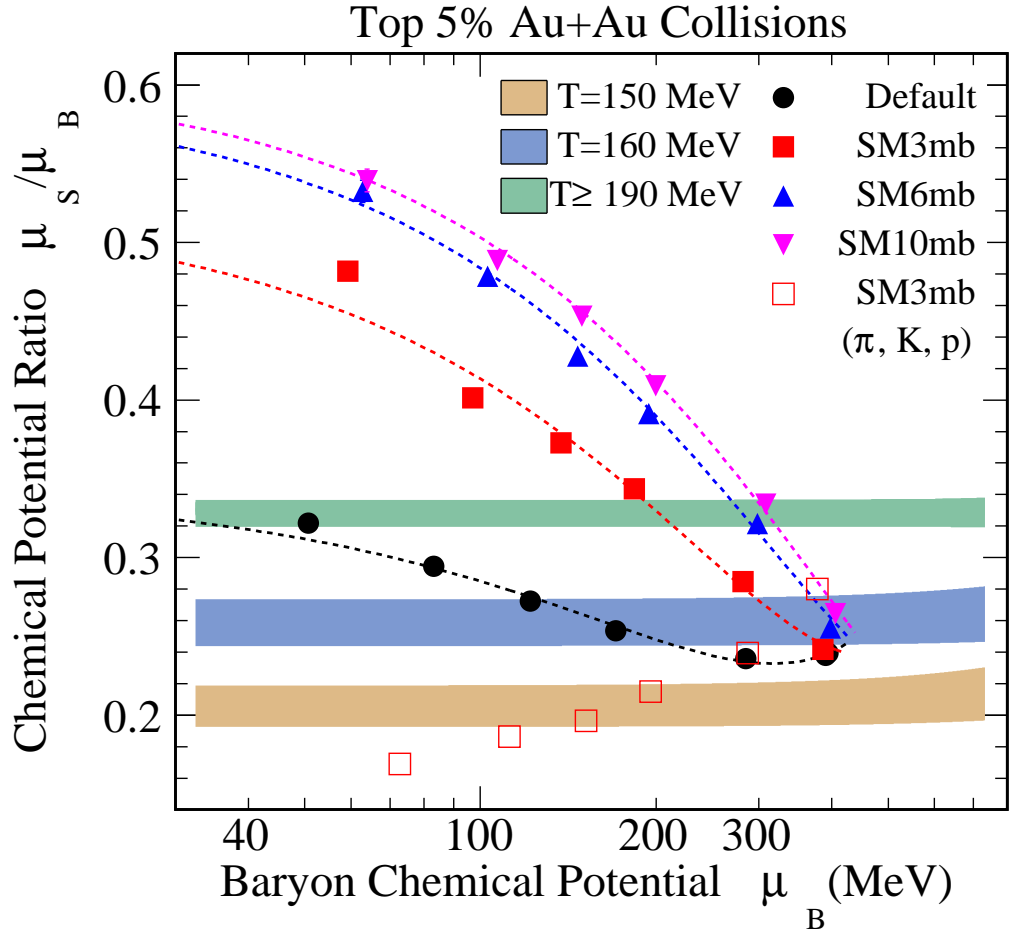}}
\caption{\label{muratios} Strangeness over baryon chemical potential ratio $\mu_S/\mu_B$ as a function of the baryon chemical potential $\mu_B$. All results are from the top 5$\%$ central Au+Au collisions at RHIC energies. In case of the SM mode, filled squares, filled triangles and filled inverted triangles are corresponding to partonic cross sections of 3, 6, and mb, respectively. The open squares are fitting results from the SM model with a parton cross section of 3 mb when only the yields of $\pi$, $K$ and $p$ are included. Results from the D model are shown as filled circles. The dashed lines are polynomial fits to the calculations. The three bands are taken from lattice QCD NLO calculations for three different critical temperature values.}
\end{figure}

Figure~\ref{otp} shows the rest of chemical freeze-out parameters from the GCE fit as functions as $N_{\text{part}}$ from the D model and SM models with 10 mb parton cross section. The behaviors of the strangeness chemical potential $\mu_S$ are similar to those of $\mu_B$. $\mu_S$ is less than $\mu_B$ for certain centralities and collision energies. The ratios of $\mu_S$ over $\mu_B$ are shown in Fig.~\ref{muratios} and compared with the results from lattice QCD calculations. The strangeness undersaturation factor $\gamma_s$, referred as the strange quark phase-space occupancy, increase with increasing energy in both the D and the SM models at a fixed collision centrality, which indicates that the deviation from strangeness chemical equilibrium becomes smaller, and the colliding systems are getting close to strangeness chemical equilibrium, with increasing energy. The deviation from strangeness chemical equilibrium is a state in which strangeness is suppressed compared to the equilibrium value. It is found that the values of $\gamma_s$ from the D and the SM models (except at some low energies) of Au+Au collisions at RHIC energies, are larger than the values of $\gamma_s\sim0.5$ reproducing the particle yields obtained in elementary collisions by canonical approach~\cite{Becattini2}. This result shows that the strangeness in these Au+Au collisions is enhanced relative to the elementary collisions. The centrality dependence of $\gamma_s$ is different between the D and the SM models at a given collision energy. In the D model, $\gamma_s$ decreases from central to peripheral, which indicates that the strangeness in central collisions is enhanced comparted to peripheral collision. In the SM model, $\gamma_s$ are nearly constant from central to peripheral and smaller than those extracted from the D model. This demonstrates that the strangeness is more under-saturated when partonic interaction is considerable, the reason for which is unknown yet. The volume $V$ of the fireball during heavy-ion collisions increase with increasing energy except 7.7 GeV from the SM model at a fixed centrality. At a given collision energy, $V$ increases linearly with increasing $N_{\text{part}}$.

At any value of ($T$, $\mu_B$), the chemical potential $\mu_S$ satisfies some constraints that can be evaluated with lattice QCD calculation~\cite{Bazavov2}. The next-to-leading order (NLO) expansion of $\mu_S/\mu_B$ can be written as 
\begin{equation}
\mu_S/\mu_B=s_1+s_3\mu_B^2
\end{equation} 
where $s_1$ and $s_3$ are two parameters which are related to the critical temperature $T$. The strangeness over baryon chemical potential ratio $\mu_S/\mu_B$ as a function of the baryon chemical potential $\mu_B$ is shown in Fig.~\ref{muratios}. These ratios are from the D model (filled-circle) and the SM models with parton cross section of 3 mb (filled square), 6 mb (filled triangle), and 10 mb (filled inverted triangle). All the results are taken from the top 5$\%$ central Au+Au collisions at RHIC energies. The dashed-lines are polynomial fits to these calculations. As a comparison, the results from lattice QCD NLO calculations for three distinctive critical temperature values, $T$=150 MeV, $T$=160 MeV and $T\geqslant$190 MeV, are also presented. The ratio $\mu_S/\mu_B$ is almost constant with the value $\frac{1}{3}$ at $T\geqslant$190 MeV from strangeness neutrality conditions in hadron resonance gas model calculations~\cite{Braun-Munzinger}. It can be found that from the D and the SM models that $\mu_B$ are nearly the same and $\mu_S/\mu_B$ ratios are different. The value of $\mu_S/\mu_B$ from the SM model is greater than that from the D model for a certain collision energy, which indicates $\mu_S$ is enhanced when partonic interaction is considered. A larger parton cross section will make the chemical potential ratio larger. The results from all the classes converge at $\mu_B\approx$ 400 MeV and $\mu_S/\mu_B$=0.24$-$0.26 which demonstrates that partonic interaction is nonsignificant in the calculation of chemical potential ratios at low collision energy, i.e., $\sqrt{s_{\text{NN}}}$=7.7 GeV. The ratios are larger at high energy collisions, especially compared to the upper limit of lattice QCD calculation, $\frac{1}{3}$, in the SM model, which is due to the unexpected antiparticle-to-particle ratios of $\bar{\Lambda}/\Lambda$, $\bar{\Xi}^+/\Xi^{-}$ and $\bar{\Omega}^+/\Omega^-$ that can be found in Fig.~\ref{pyield}. The relations between $\mu_S$ and these ratios are

\begin{eqnarray}
\frac{\bar{\Lambda}}{\Lambda}&=&\text{exp}\left( \frac{-2\mu_B}{T}+\frac{2\mu_S}{T}\right)\nonumber\\
\frac{\bar{\Xi}^+}{\Xi^-}&=&\text{exp}\left( \frac{-2\mu_B}{T}+\frac{4\mu_S}{T}\right)\\
\frac{\bar{\Omega}^+}{\Omega^-}&=&\text{exp}\left( \frac{-2\mu_B}{T}+\frac{6\mu_S}{T}\right)\nonumber
\end{eqnarray}

where $\text{exp}\left( \frac{-2\mu_B}{T}\right) =\dfrac{\bar{p}}{p}$. These unexpected ratios in the SM model can lead to large $\mu_S$ and $\mu_S/\mu_B$. The $\mu_S/\mu_B$ ratios extracted from the SM model with parton cross section of 3 mb (open squares) when only the yields of $\pi$, $K$ and $p$ are included in the THERMUS fit are also shown in Fig.~\ref{muratios}. It can be concluded that the strange baryon antiparticle-to-particle ratios, which are anomalous in the SM model, are important for extracting the $\mu_S$ and $\mu_S/\mu_B$ ratios.
 
In summary, by using the AMPT and the THERMUS models, we perform a systematic analysis of the chemical freeze-out parameters of high energy heavy-ion collision at RHIC energy in order to provide a reference for the BES program. It was found that at a fixed collision centrality, the temperature $T_{\text{ch}}$ has weak dependence on energy in the D model; however, it increases with energy and reaches saturation near 10 GeV in the SM model. The partonic interaction is essential for determining the yields of protons(antiprotons) and the value of this saturation temperature. At any given collision energy, $T_{\text{ch}}$ decreases from central to peripheral in all cases. The chemical potentials of baryons $\mu_B$, and strangeness $\mu_S$ decrease with energy at a fixed centrality and decrease from central to peripheral at a given collision energy. The partonic interaction makes significant changes to $T_{\text{ch}}$ and insignificant changes to $\mu_B$. The strangeness undersaturation factor $\gamma_S$, parameterized the degree of equilibration, increases with energy at fixed centrality, which indicates the matter created in heavy-ion collisions is getting close to strangeness chemical equilibrium with increasing energy. At a given collision energy, $\gamma_S$ decreases from central to peripheral in the D model and is almost constant in the SM model. The strangeness is more undersaturated when the partonic interaction is considerable. The volume of the fireball increases with energy and has a linear relation to $N_{\text{part}}$. The chemical potential ratios $\mu_S/\mu_B$ of all different classes of AMPT models converge at lowest energy 7.7 GeV. The reason that these ratios are larger than the upper limit of the lattice QCD calculation in the SM model is due to the unreasonable ratios of multistrange antiparticles to particles. Overall, the D model describes both the particle yields and the chemical freeze-out parameters better than the SM model. 

We thank F. Karsch and N. Xu for discussion and help with the chemical potential ratios from lattice QCD. This work is supported in part by China Postdoctoral Science Foundation No.2013M542040, the National Natural Science Foundation of China under Grants No.11228513, No.11221504, No.11135011 and No.11247263.

\bibliography{apssamp}

\providecommand{\noopsort}[1]{}\providecommand{\singleletter}[1]{#1}%
\begin{thebibliography}{23}%
\makeatletter
\providecommand \@ifxundefined [1]{%
 \@ifx{#1\undefined}
}%
\providecommand \@ifnum [1]{%
 \ifnum #1\expandafter \@firstoftwo
 \else \expandafter \@secondoftwo
 \fi
}%
\providecommand \@ifx [1]{%
 \ifx #1\expandafter \@firstoftwo
 \else \expandafter \@secondoftwo
 \fi
}%
\providecommand \natexlab [1]{#1}%
\providecommand \enquote  [1]{``#1''}%
\providecommand \bibnamefont  [1]{#1}%
\providecommand \bibfnamefont [1]{#1}%
\providecommand \citenamefont [1]{#1}%
\providecommand \href@noop [0]{\@secondoftwo}%
\providecommand \href [0]{\begingroup \@sanitize@url \@href}%
\providecommand \@href[1]{\@@startlink{#1}\@@href}%
\providecommand \@@href[1]{\endgroup#1\@@endlink}%
\providecommand \@sanitize@url [0]{\catcode `\\12\catcode `\$12\catcode
  `\&12\catcode `\#12\catcode `\^12\catcode `\_12\catcode `\%12\relax}%
\providecommand \@@startlink[1]{}%
\providecommand \@@endlink[0]{}%
\providecommand \url  [0]{\begingroup\@sanitize@url \@url }%
\providecommand \@url [1]{\endgroup\@href {#1}{\urlprefix }}%
\providecommand \urlprefix  [0]{URL }%
\providecommand \Eprint [0]{\href }%
\providecommand \doibase [0]{http://dx.doi.org/}%
\providecommand \selectlanguage [0]{\@gobble}%
\providecommand \bibinfo  [0]{\@secondoftwo}%
\providecommand \bibfield  [0]{\@secondoftwo}%
\providecommand \translation [1]{[#1]}%
\providecommand \BibitemOpen [0]{}%
\providecommand \bibitemStop [0]{}%
\providecommand \bibitemNoStop [0]{.\EOS\space}%
\providecommand \EOS [0]{\spacefactor3000\relax}%
\providecommand \BibitemShut  [1]{\csname bibitem#1\endcsname}%
\let\auto@bib@innerbib\@empty
\bibitem [{\citenamefont {Adams}\ and\ \citenamefont {\textit{et al.}
  (STAR~Collaboration)}(2005{\natexlab{a}})}]{Star1-1}%
  \BibitemOpen
  \bibfield  {author} {\bibinfo {author} {\bibfnamefont {J.}~\bibnamefont
  {Adams}}\ and\ \bibinfo {author} {\bibnamefont {\textit{et al.}
  (STAR~Collaboration)}},\ }\href@noop {} {\bibfield  {journal} {\bibinfo
  {journal} {Nucl. Phys.}\ }\textbf {\bibinfo {volume} {A757}},\ \bibinfo
  {pages} {102} (\bibinfo {year} {2005}{\natexlab{a}})}\BibitemShut {NoStop}%
\bibitem [{\citenamefont {Stephanov}(2004{\natexlab{a}})}]{Stephanov-1}%
  \BibitemOpen
  \bibfield  {author} {\bibinfo {author} {\bibfnamefont {M.~A.}\ \bibnamefont
  {Stephanov}},\ }\href@noop {} {\bibfield  {journal} {\bibinfo  {journal}
  {Prog. Theor. Phys. Suppl.}\ }\textbf {\bibinfo {volume} {153}},\ \bibinfo
  {pages} {139} (\bibinfo {year} {2004}{\natexlab{a}})}\BibitemShut {NoStop}%
\bibitem [{\citenamefont {Stephanov}(2004{\natexlab{b}})}]{Stephanov-2}%
  \BibitemOpen
  \bibfield  {author} {\bibinfo {author} {\bibfnamefont {M.~A.}\ \bibnamefont
  {Stephanov}},\ }\href@noop {} {} (\bibinfo {year} {2004}{\natexlab{b}}),\
  \Eprint {http://arxiv.org/abs/hep-ph/0402115} {hep-ph/0402115} \BibitemShut
  {NoStop}%
\bibitem [{\citenamefont {Mohanty}(2009)}]{Mohanty}%
  \BibitemOpen
  \bibfield  {author} {\bibinfo {author} {\bibfnamefont {B.}~\bibnamefont
  {Mohanty}},\ }\href@noop {} {\bibfield  {journal} {\bibinfo  {journal} {Nucl.
  Phys.}\ }\textbf {\bibinfo {volume} {A830}},\ \bibinfo {pages} {899c}
  (\bibinfo {year} {2009})}\BibitemShut {NoStop}%
\bibitem [{\citenamefont {Cleymans}\ \emph {et~al.}(2002)\citenamefont
  {Cleymans}, \citenamefont {Kampfer},\ and\ \citenamefont
  {Wheaton}}]{Cleymans1}%
  \BibitemOpen
  \bibfield  {author} {\bibinfo {author} {\bibfnamefont {J.}~\bibnamefont
  {Cleymans}}, \bibinfo {author} {\bibfnamefont {B.}~\bibnamefont {Kampfer}}, \
  and\ \bibinfo {author} {\bibfnamefont {S.}~\bibnamefont {Wheaton}},\
  }\href@noop {} {\bibfield  {journal} {\bibinfo  {journal} {Phys. Rev. C}\
  }\textbf {\bibinfo {volume} {65}},\ \bibinfo {pages} {027901} (\bibinfo
  {year} {2002})}\BibitemShut {NoStop}%
\bibitem [{\citenamefont {Becattini}\ \emph {et~al.}(2006)\citenamefont
  {Becattini}, \citenamefont {Manninen},\ and\ \citenamefont
  {Gazdzicki}}]{Becattini1}%
  \BibitemOpen
  \bibfield  {author} {\bibinfo {author} {\bibfnamefont {F.}~\bibnamefont
  {Becattini}}, \bibinfo {author} {\bibfnamefont {J.}~\bibnamefont {Manninen}},
  \ and\ \bibinfo {author} {\bibfnamefont {M.}~\bibnamefont {Gazdzicki}},\
  }\href@noop {} {\bibfield  {journal} {\bibinfo  {journal} {Phys. Rev. C}\
  }\textbf {\bibinfo {volume} {73}},\ \bibinfo {pages} {044905} (\bibinfo
  {year} {2006})}\BibitemShut {NoStop}%
\bibitem [{\citenamefont {Vogt}(2007)}]{Vogt}%
  \BibitemOpen
  \bibfield  {author} {\bibinfo {author} {\bibfnamefont {R.}~\bibnamefont
  {Vogt}},\ }\href@noop {} {\emph {\bibinfo {title} {Ultrarelativistic
  Heavy-Ion Collisions}}}\ (\bibinfo  {publisher} {Elsevier Science Ltd},\
  \bibinfo {year} {2007})\BibitemShut {NoStop}%
\bibitem [{\citenamefont {Fodor}\ and\ \citenamefont {Katz}(2009)}]{Fodor}%
  \BibitemOpen
  \bibfield  {author} {\bibinfo {author} {\bibfnamefont {Z.}~\bibnamefont
  {Fodor}}\ and\ \bibinfo {author} {\bibfnamefont {S.~D.}\ \bibnamefont
  {Katz}},\ }\href@noop {} {} (\bibinfo {year} {2009}),\ \Eprint
  {http://arxiv.org/abs/0908.3341} {0908.3341} \BibitemShut {NoStop}%
\bibitem [{\citenamefont {Bazavov}\ and\ \citenamefont {\textit{et
  al.}}(2009)}]{Bazavov1}%
  \BibitemOpen
  \bibfield  {author} {\bibinfo {author} {\bibfnamefont {A.}~\bibnamefont
  {Bazavov}}\ and\ \bibinfo {author} {\bibnamefont {\textit{et al.}}},\
  }\href@noop {} {\bibfield  {journal} {\bibinfo  {journal} {Phys. Rev. D}\
  }\textbf {\bibinfo {volume} {80}},\ \bibinfo {pages} {014504} (\bibinfo
  {year} {2009})}\BibitemShut {NoStop}%
\bibitem [{\citenamefont {Wang}\ and\ \citenamefont {Gyulassy}(1991)}]{HIJING}%
  \BibitemOpen
  \bibfield  {author} {\bibinfo {author} {\bibfnamefont {X.~N.}\ \bibnamefont
  {Wang}}\ and\ \bibinfo {author} {\bibfnamefont {M.}~\bibnamefont
  {Gyulassy}},\ }\href@noop {} {\bibfield  {journal} {\bibinfo  {journal}
  {Phys. Rev. D}\ }\textbf {\bibinfo {volume} {44}},\ \bibinfo {pages} {3501}
  (\bibinfo {year} {1991})}\BibitemShut {NoStop}%
\bibitem [{\citenamefont {Zhang}(1998)}]{ZPC}%
  \BibitemOpen
  \bibfield  {author} {\bibinfo {author} {\bibfnamefont {B.}~\bibnamefont
  {Zhang}},\ }\href@noop {} {\bibfield  {journal} {\bibinfo  {journal} {Comput.
  Phys. Commun.}\ }\textbf {\bibinfo {volume} {109}},\ \bibinfo {pages} {193}
  (\bibinfo {year} {1998})}\BibitemShut {NoStop}%
\bibitem [{\citenamefont {Andersson}\ \emph {et~al.}(1983)\citenamefont
  {Andersson}, \citenamefont {Gustafson}, \citenamefont {Ingelman},\ and\
  \citenamefont {Sjostrand}}]{Lund}%
  \BibitemOpen
  \bibfield  {author} {\bibinfo {author} {\bibfnamefont {B.}~\bibnamefont
  {Andersson}}, \bibinfo {author} {\bibfnamefont {G.}~\bibnamefont
  {Gustafson}}, \bibinfo {author} {\bibfnamefont {G.}~\bibnamefont {Ingelman}},
  \ and\ \bibinfo {author} {\bibfnamefont {T.}~\bibnamefont {Sjostrand}},\
  }\href@noop {} {\bibfield  {journal} {\bibinfo  {journal} {Phys. Rep}\
  }\textbf {\bibinfo {volume} {97}},\ \bibinfo {pages} {31} (\bibinfo {year}
  {1983})}\BibitemShut {NoStop}%
\bibitem [{\citenamefont {Li}\ and\ \citenamefont {Ko}(1995)}]{ART}%
  \BibitemOpen
  \bibfield  {author} {\bibinfo {author} {\bibfnamefont {B.~A.}\ \bibnamefont
  {Li}}\ and\ \bibinfo {author} {\bibfnamefont {C.~M.}\ \bibnamefont {Ko}},\
  }\href@noop {} {\bibfield  {journal} {\bibinfo  {journal} {Phys. Rev. C}\
  }\textbf {\bibinfo {volume} {52}},\ \bibinfo {pages} {2037} (\bibinfo {year}
  {1995})}\BibitemShut {NoStop}%
\bibitem [{\citenamefont {Wheaton}\ \emph {et~al.}(2009)\citenamefont
  {Wheaton}, \citenamefont {Cleymans},\ and\ \citenamefont {Hauer}}]{THERMUS}%
  \BibitemOpen
  \bibfield  {author} {\bibinfo {author} {\bibfnamefont {S.}~\bibnamefont
  {Wheaton}}, \bibinfo {author} {\bibfnamefont {J.}~\bibnamefont {Cleymans}}, \
  and\ \bibinfo {author} {\bibfnamefont {M.}~\bibnamefont {Hauer}},\
  }\href@noop {} {\bibfield  {journal} {\bibinfo  {journal} {Comp. Phys.
  Commun.}\ }\textbf {\bibinfo {volume} {180}},\ \bibinfo {pages} {84}
  (\bibinfo {year} {2009})}\BibitemShut {NoStop}%
\bibitem [{\citenamefont {Andronic}\ \emph {et~al.}(2010)\citenamefont
  {Andronic}, \citenamefont {Braun-Munzinger},\ and\ \citenamefont
  {Stachel}}]{Andronic}%
  \BibitemOpen
  \bibfield  {author} {\bibinfo {author} {\bibfnamefont {A.}~\bibnamefont
  {Andronic}}, \bibinfo {author} {\bibfnamefont {P.}~\bibnamefont
  {Braun-Munzinger}}, \ and\ \bibinfo {author} {\bibfnamefont {J.}~\bibnamefont
  {Stachel}},\ }\href@noop {} {\bibfield  {journal} {\bibinfo  {journal} {Nucl.
  Phys. A}\ }\textbf {\bibinfo {volume} {834}},\ \bibinfo {pages} {237c}
  (\bibinfo {year} {2010})}\BibitemShut {NoStop}%
\bibitem [{\citenamefont {Cleymans}\ \emph {et~al.}(2006)\citenamefont
  {Cleymans}, \citenamefont {Oeschler}, \citenamefont {Redlich},\ and\
  \citenamefont {Wheaton}}]{Cleymans2}%
  \BibitemOpen
  \bibfield  {author} {\bibinfo {author} {\bibfnamefont {J.}~\bibnamefont
  {Cleymans}}, \bibinfo {author} {\bibfnamefont {H.}~\bibnamefont {Oeschler}},
  \bibinfo {author} {\bibfnamefont {K.}~\bibnamefont {Redlich}}, \ and\
  \bibinfo {author} {\bibfnamefont {S.}~\bibnamefont {Wheaton}},\ }\href@noop
  {} {\bibfield  {journal} {\bibinfo  {journal} {Phys. Rev. C}\ }\textbf
  {\bibinfo {volume} {73}},\ \bibinfo {pages} {034905} (\bibinfo {year}
  {2006})}\BibitemShut {NoStop}%
\bibitem [{\citenamefont {Abelev}\ and\ \citenamefont {\textit{et al.}
  (STAR~Collaboration)}(2009)}]{Star3}%
  \BibitemOpen
  \bibfield  {author} {\bibinfo {author} {\bibfnamefont {B.~I.}\ \bibnamefont
  {Abelev}}\ and\ \bibinfo {author} {\bibnamefont {\textit{et al.}
  (STAR~Collaboration)}},\ }\href@noop {} {\bibfield  {journal} {\bibinfo
  {journal} {Phys. Rev. C}\ }\textbf {\bibinfo {volume} {79}},\ \bibinfo
  {pages} {034909} (\bibinfo {year} {2009})}\BibitemShut {NoStop}%
\bibitem [{\citenamefont {Adams}\ and\ \citenamefont {\textit{et al.}
  (STAR~Collaboration)}(2007)}]{Star4}%
  \BibitemOpen
  \bibfield  {author} {\bibinfo {author} {\bibfnamefont {J.}~\bibnamefont
  {Adams}}\ and\ \bibinfo {author} {\bibnamefont {\textit{et al.}
  (STAR~Collaboration)}},\ }\href@noop {} {\bibfield  {journal} {\bibinfo
  {journal} {Phys. Rev. Lett.}\ }\textbf {\bibinfo {volume} {98}},\ \bibinfo
  {pages} {062301} (\bibinfo {year} {2007})}\BibitemShut {NoStop}%
\bibitem [{\citenamefont {Adams}\ and\ \citenamefont {\textit{et al.}
  (STAR~Collaboration)}(2005{\natexlab{b}})}]{Star5}%
  \BibitemOpen
  \bibfield  {author} {\bibinfo {author} {\bibfnamefont {J.}~\bibnamefont
  {Adams}}\ and\ \bibinfo {author} {\bibnamefont {\textit{et al.}
  (STAR~Collaboration)}},\ }\href@noop {} {\bibfield  {journal} {\bibinfo
  {journal} {Phys. Lett.}\ }\textbf {\bibinfo {volume} {B612}},\ \bibinfo
  {pages} {181} (\bibinfo {year} {2005}{\natexlab{b}})}\BibitemShut {NoStop}%
\bibitem [{\citenamefont {Cleymans}\ \emph {et~al.}(2005)\citenamefont
  {Cleymans}, \citenamefont {Kampfer}, \citenamefont {Kaneta}, \citenamefont
  {Wheaton},\ and\ \citenamefont {Xu}}]{Cleymans3}%
  \BibitemOpen
  \bibfield  {author} {\bibinfo {author} {\bibfnamefont {J.}~\bibnamefont
  {Cleymans}}, \bibinfo {author} {\bibfnamefont {B.}~\bibnamefont {Kampfer}},
  \bibinfo {author} {\bibfnamefont {M.}~\bibnamefont {Kaneta}}, \bibinfo
  {author} {\bibfnamefont {S.}~\bibnamefont {Wheaton}}, \ and\ \bibinfo
  {author} {\bibfnamefont {N.}~\bibnamefont {Xu}},\ }\href@noop {} {\bibfield
  {journal} {\bibinfo  {journal} {Phys. Rev. C}\ }\textbf {\bibinfo {volume}
  {71}},\ \bibinfo {pages} {054901} (\bibinfo {year} {2005})}\BibitemShut
  {NoStop}%
\bibitem [{\citenamefont {Becattini}\ and\ \citenamefont
  {Heinz}(1997)}]{Becattini2}%
  \BibitemOpen
  \bibfield  {author} {\bibinfo {author} {\bibfnamefont {F.}~\bibnamefont
  {Becattini}}\ and\ \bibinfo {author} {\bibfnamefont {U.}~\bibnamefont
  {Heinz}},\ }\href@noop {} {\bibfield  {journal} {\bibinfo  {journal} {Z.
  Phys. C}\ }\textbf {\bibinfo {volume} {76}},\ \bibinfo {pages} {269}
  (\bibinfo {year} {1997})}\BibitemShut {NoStop}%
\bibitem [{\citenamefont {Bazavov}\ and\ \citenamefont {\textit{et
  al.}}(2012)}]{Bazavov2}%
  \BibitemOpen
  \bibfield  {author} {\bibinfo {author} {\bibfnamefont {A.}~\bibnamefont
  {Bazavov}}\ and\ \bibinfo {author} {\bibnamefont {\textit{et al.}}},\
  }\href@noop {} {\bibfield  {journal} {\bibinfo  {journal} {Phys. Rev. Lett.}\
  }\textbf {\bibinfo {volume} {109}},\ \bibinfo {pages} {192302} (\bibinfo
  {year} {2012})}\BibitemShut {NoStop}%
\bibitem [{\citenamefont {Braun-Munzinger}\ \emph {et~al.}(2003)\citenamefont
  {Braun-Munzinger}, \citenamefont {Redlich},\ and\ \citenamefont
  {Stachel}}]{Braun-Munzinger}%
  \BibitemOpen
  \bibfield  {author} {\bibinfo {author} {\bibfnamefont {P.}~\bibnamefont
  {Braun-Munzinger}}, \bibinfo {author} {\bibfnamefont {K.}~\bibnamefont
  {Redlich}}, \ and\ \bibinfo {author} {\bibfnamefont {J.}~\bibnamefont
  {Stachel}},\ }\href@noop {} {} (\bibinfo {year} {2003}),\ \Eprint
  {http://arxiv.org/abs/nucl-th/0304013} {nucl-th/0304013} \BibitemShut
  {NoStop}%
\end{thebibliography}%

\end{document}